\documentclass{emulateapj}
 
\shorttitle{WASP-47}
\shortauthors{Dai et al.}

\usepackage{amsmath}
\usepackage{epstopdf}
\usepackage{apjfonts}
\usepackage{xspace}
\usepackage{hyperref} 
\usepackage{bm}
\usepackage{graphicx}
\begin{document}


\title{Doppler Monitoring of the WASP-47 multiplanet
  system$^\star$}


\author{Fei\ Dai\altaffilmark{1}, Joshua N.\ Winn\altaffilmark{1}, Pamela\ Arriagada\altaffilmark{2}, R.Paul\ Butler\altaffilmark{2}, Jeffrey D.\ Crane\altaffilmark{3}, John Asher\ Johnson\altaffilmark{4},\\ Stephen A.\ Shectman\altaffilmark{3}, Johanna K.\ Teske\altaffilmark{2,3}, Ian B.\ Thompson\altaffilmark{3}, Andrew\ Vanderburg\altaffilmark{4}, Robert A.\ Wittenmyer\altaffilmark{5,6}}


\altaffiltext{$\star$}{This paper includes data gathered with the 6.5 meter Magellan Telescopes located at Las Campanas Observatory, Chile.}
\altaffiltext{1}{Department of Physics and Kavli Institute for Astrophysics and Space Research,
  Massachusetts Institute of Technology, Cambridge, MA 02139, USA {\tt
    fd284@mit.edu}}
\altaffiltext{2}{Carnegie Institution of Washington, Department of Terrestrial Magnetism, 5241 Broad Branch Road, NW, Washington DC, 20015-1305, USA}
\altaffiltext{3}{The Observatories of the Carnegie Institution of Washington, 813 Santa Barbara Street, Pasadena, CA 91101, USA}
\altaffiltext{4}{Harvard-Smithsonian Center for Astrophysics, 60 Garden Street, Cambridge, MA 02138, USA }
\altaffiltext{5}{Exoplanetary Science at UNSW, School of Physics, UNSW Australia, Sydney, NSW 2052, Australia}
\altaffiltext{6}{Australian Centre for Astrobiology, UNSW Australia, Sydney, NSW 2052, Australia}


\begin{abstract}
\noindent
We present precise Doppler observations of WASP-47, a transiting
planetary system featuring a hot Jupiter with both inner and outer
planetary companions. This system has an unusual architecture and also
provides a rare opportunity to measure planet masses in two different
ways: the Doppler method, and the analysis of transit-timing
variations (TTV).  Based on the new Doppler data, obtained with the
Planet Finder Spectrograph on the Magellan/Clay 6.5m telescope, the
mass of the hot Jupiter is $370 \pm 29~M_{\oplus}$. This is consistent
with the previous Doppler determination as well as the TTV
determination.  For the inner planet WASP-47e, the Doppler data lead
to a mass of $12.2\pm 3.7~ M_{\oplus}$, in agreement with the
TTV-based upper limit of $<$22~$M_{\oplus}$ ($95\%$ confidence). For
the outer planet WASP-47d, the Doppler mass constraint of $10.4\pm
8.4~M_{\oplus}$ is consistent with the TTV-based measurement of
$15.2^{+6.7}_{-7.6}~ M_{\oplus}$.
\end{abstract}


\section{Introduction}

The first two things one wants to know about any newly discovered
planet are its mass and radius. Is it relatively small and dense,
similar to Earth?  Is it large and diffuse, similar to Jupiter and
Saturn?  Or is it somewhere in between? Although the {\it Kepler}
mission revolutionized exoplanetary science, for the specific purpose
of measuring planet masses the original mission was not ideal. This is
because the typical target stars were relatively faint ($V=$~14-16),
frustrating efforts to obtain high-resolution spectra that are
necessary to measure planetary masses by the Doppler method. For
planets smaller than Neptune, it has only been possible to measure the
masses of a few dozen {\it Kepler} planets with the brightest host
stars \citep{marcy2014,Howard2013,Pepe2013,Dressing2015}, and even in
those cases many of the mass measurements have large uncertainties.

Because of a mechanical failure and reduction in capability, the {\it
  Kepler} telescope abandoned its original mission and is now engaged
in a new mission called {\it K2} \citep{howell}. Every 3 months, the
telescope observes a different field on the ecliptic (the only zone
where it can achieve stable pointing), providing a fresh sample of
bright stars for which precise Doppler observations are possible. The
third {\it K2} field happened to encompass WASP-47, a G9 star with a
previously discovered transiting hot Jupiter \citep{Hellier}. The {\it
  K2} data have revealed two additional transiting planets, with
periods of 0.8 days and 9 days \citep{Becker}. Furthermore,
\citet{Neveu} recently reported the Doppler discovery of a second,
wide-orbiting Jovian planet, with a period of $572 \pm 7$~days.

With this discovery, WASP-47 is unique among the known exoplanetary
systems: a hot Jupiter with very close companions on both interior and
exterior orbits, in addition to a distant companion. The close
companions have implications for theories of hot Jupiter formation. It
seems difficult to attain such a compact and fragile configuration
within violent scenarios such as planet-planet scattering
\citep{Rasio,Weidenschilling} or Kozai-Lidov oscillations
\citep{Mazeh, Holman1997, Innanen}. The WASP-47 system also presents a
rare opportunity to measure the masses of the inner three planets
using two independent techniques: the traditional Doppler method, and
the analysis of transit-timing variations
\citep[TTV,][]{Holmanmurray,Agol2005}. It is always useful to have
independent methods for measuring important quantities, and for planet
masses in particular, there is some controversy over the reliability
of TTV-based masses. The TTV method has revealed some planets with
surprisingly low densities \citep[see,
e.g.,][]{Lissauer2013,Schmitt,Jontof2014,Jontof2015}. Among the sample
of planets smaller than $4.0~R_{\oplus}$, those for which masses have
been determined with TTVs have systematically lower masses than the
subsample for which masses have been determined with the Doppler
method \citep{weissmarcy2014}. This discrepancy could be due to
systematic errors in one or both methods, along with biases in the
various samples. To disentangle these effects, it is useful to
identify individual systems for which the Doppler and TTV methods can
both be applied.

\begin{deluxetable}{lll}
\tabletypesize{\scriptsize}
\tablecaption{Key parameters of the WASP-47 system\label{tbl:params}}
\tablewidth{0pt}

\tablehead{
\colhead{Parameter} & \colhead{Value and 68.3\% Conf.~Limits} & \colhead{Ref.} 
}
\startdata

WASP-47b\\
Orbital period [days]  & $4.1591282 \pm 0.0000046 $  & A\\
Transit epoch [BJD]  & $2457007.932132 \pm 0.000061 $  & A\\
Radius [$R_{\oplus}$] & $12.71 \pm 0.44 $  & A\\
RV semi-amplitude $K$ [m~$s^{-1}$] & $143.3 \pm 2.5$  & B\\
RV-based mass [$M_{\oplus}$] & $370 \pm 29$  & B\\
TTV-based mass [$M_{\oplus}$] & $341^{+73}_{-55}$  & A\\
Mean density (RV-based) [g~cm$^{-3}$] & $0.99 \pm 0.13$  & A, B\\
  \\
WASP-47e\\
Orbital period [days] & $0.789593 \pm 0.000012 $  & A\\
Transit epoch [BJD] & $2457011.34859 \pm 0.00031 $  & A\\
Radius [$R_{\oplus}$] & $1.817 \pm 0.065 $  & A\\
RV semi-amplitude $K$ [m~s$^{-1}$] & $8.2 \pm 2.4$  & B\\
RV-based mass [$M_{\oplus}$] & $12.2\pm 3.7$  & B\\
TTV-based mass [$M_{\oplus}$] & $<22$ ($95\%$ Confidence)  & A\\
Mean density (RV-based) [g~cm$^{-3}$] & $11.2 \pm 3.6$  & A, B\\
  \\
WASP-47d\\
Orbital period [days] & $9.03079 \pm 0.00017 $  & A\\
Transit epoch [BJD] & $2457006.36922 \pm 0.00052 $  & A\\
Radius [$R_{\oplus}$] & $3.60 \pm 0.13 $  & A\\
RV semi-amplitude $K$ [m~s$^{-1}$] & $3.1 \pm 2.5$  & B\\
RV-based mass [$M_{\oplus}$] & $10.4\pm 8.4$  & B\\
TTV-based mass [$M_{\oplus}$] & $15.2^{+6.7}_{-7.6}$  & A\\
Mean density (RV-based) [g~cm$^{-3}$] & $1.2 \pm 1.0$  & A, B\\
\enddata
\tablecomments{A: \citet{Becker}; B: This work.}
\end{deluxetable}

\begin{table}
\centering
\caption{Relative radial velocity of WASP-47}
\begin{tabular}{lrr}
\hline
\hline
{\rm BJD}  & {\rm RV [m~s$^{-1}$]} & {\rm Unc.\ [m~s$^{-1}$]}  \\
\hline
 2457257.721181 &$  -70.6 $&$  2.5$\\
 2457257.751794 &$  -77.9 $&$  2.7$\\
 2457257.783981 &$  -91.4 $&$  2.8$\\
 2457257.821563 &$  -97.1 $&$  2.9$\\
 2457258.718380 &$ -162.9 $&$  3.4$\\
 2457258.790174 &$ -154.4 $&$  2.7$\\
 2457258.857569 &$ -132.2 $&$  3.5$\\
 2457261.736597 &$  -49.8 $&$  2.7$\\
 2457261.767025 &$  -56.8 $&$  3.4$\\
 2457261.821400 &$  -61.7 $&$  4.0$\\
 2457264.621262 &$  130.9 $&$  3.1$\\
 2457264.722280 &$  137.2 $&$  3.3$\\
 2457264.762627 &$  135.0 $&$  3.9$\\
 2457264.792650 &$  137.5 $&$  3.9$\\
 2457267.665532 &$  -56.2 $&$  2.8$\\
 2457267.718310 &$  -32.4 $&$  2.7$\\
 2457267.771910 &$  -42.8 $&$  2.7$\\
 2457267.816458 &$  -18.3 $&$  3.4$\\
 2457268.571817 &$  112.6 $&$  3.3$\\
 2457268.673646 &$  118.0 $&$  3.1$\\
 2457268.760174 &$  122.2 $&$  3.4$\\
 2457268.798935 &$  113.0 $&$  3.9$\\
 2457269.608519 &$   39.0 $&$  2.8$\\
 2457269.710486 &$   27.7 $&$  3.1$\\
 2457269.747708 &$   27.7 $&$  2.7$\\
 2457269.793981 &$    3.4 $&$  3.7$\\
\hline
\end{tabular}
\label{rv_table}
\end{table} 

WASP-47 is precisely such an example. \citet{Becker} have already
performed a TTV analysis of the available {\it K2} data for
WASP-47. In this Letter we present new Doppler observations that have
led to complementary mass determinations. The new data are described
in Section \ref{sec:observations}, our analysis is presented in
Section \ref{sec:analysis}, and the implications are discussed in
Section \ref{sec:discussion}.
 \\
 \\
\section{Observations}
\label{sec:observations}

WASP-47 was observed from August 23rd to September 4th 2015 UT with
the Carnegie Planet Finder Spectrograph \citep[PFS,][]{crane} on the
6.5 meter Magellan/Clay Telescope at Las Campanas Observatory in
Chile. We obtained several spectra of WASP-47 on each clear night, for
a total of 27 spectra. We employed an iodine gas cell to superimpose
well-characterized absorption features onto the stellar spectrum,
helping to establish the wavelength scale and instrumental profile.
The detector was read out in the $2\times 2$ binned mode, to reduce
readout noise. The typical exposure time was avout 20 minutes, giving
a signal-to-noise ratio of $\approx$73~pixel$^{-1}$ and a resolution
of about 76000 in the vicinity of the iodine absorption lines. An
additional spectrum with higher resolution and signal-to-noise ratio
was obtained without the iodine cell, to serve as a template spectrum
for the Doppler analysis.

The relative radial velocities were extracted from the spectrum using
the techniques of \citet{Butler}. The internal measurement
uncertainties (ranging from 2.5-4~m~s$^{-1}$) were estimated from the
scatter in the results to fitting individual 2~\AA~sections of the
spectrum. Table \ref{rv_table} gives the radial velocities and the internal
measurement uncertanties. Figure \ref{rvall} shows the observed radial
velocities.

\section{Analysis}
\label{sec:analysis}

We will refer to the WASP-47 planets as follows: planet b is the
transiting hot Jupiter; planet c is the long-period Jupiter identified
by \citet{Neveu}; planet d is the transiting planet with period
9.0~days and planet e is the transiting planet with period 0.8~days.

Before modeling the Doppler data, an important question is whether it
suffices to model the stellar motion as the superposition of
non-interacting orbits, or whether the gravitational interactions
between planets need be taken into account (which requires far more
computation time). We concluded that gravitational interactions could
be neglected for present purposes, based on the following test. First,
we fitted the data with a model consisting of four non-interacting
Keplerian orbits. Then, we took the best-fitting model parameters as
the starting conditions for a dynamical five-body integration of
Newton's equations of motion, using the 4th order Hermite scheme
that is available on the {\it Systemic} console \citep{systemic}. We
examined the deviations between the RVs calculated in the dynamical
model and the RVs in the non-interacting model. Over the relatively
short timespan of our PFS observations, the maximum deviation is only
0.14~m~s$^{-1}$, which is much smaller than the uncertainties in the RV
data and the uncertainties in the amplitudes of the Doppler signals.

We assumed the orbits of the inner three planets to be circular
because tidal circularization timescales are expected to be short, at
least for the two inner planets.  Furthermore, \citet{Becker} showed
that low eccentricities ($e<0.05$) are required for all three planets
in order to ensure long-term dynamical stability of the system. And
from a practical point of view, with only 26 data points and
relatively small Doppler amplitudes, we can gain little empirical
information at this stage regarding orbital eccentricities.

Given the long period of planet c ($572 \pm 7$ days), and the
relatively short interval of our PFS observations, we do not attempt
to characterize planet c. Instead, we adopted the best-fitting
parameters for planet c reported by \citet{Neveu}, and subtracted its
expected contribution from the Doppler data, prior to our
analysis. (Although we performed this step for completeness, in
practice the contribution from planet c has no substantial effect on
the results.)

With these choices, our model has 5 free parameters: the
semi-amplitude $K$ of the radial-velocity variation induced by each of
the 3 inner planets, an arbitrary additive constant $\gamma$ (since
only the relative radial velocities are measured precisely), and a
``jitter'' term $\sigma_j$ that is added in quadrature with the
internal measurement uncertainty. The jitter term is intended to
account for additional sources of uncorrelated uncertainties, which
could be of astrophysical or instrumental origin. We held fixed the
orbital periods and transit epochs at the values reported by
\citet{Becker}, as they have negligible uncertainties for our
purposes.  We adopted a likelihood function
\begin{equation}
\mathcal{L}= \prod_{i=1}^{N}\left({\frac{1}{\sqrt{2 \pi (\sigma_i^2 + \sigma_{sj}^2)}} \exp \left[ - \frac{[RV(t_i) - \mathcal{M}(t_i)]^2}{2 (\sigma_i^2+\sigma_j^2)}\right] }\right),
\end{equation}
where $RV(t_i)$ is the measured radial velocity
at time $t_i$; $\mathcal{M}(t_i)$ is the calculated radial velocity
at time $t_i$ for a particular choice of model parameters;
$\sigma_{i}$ is the internal measurement uncertainty;
and $\sigma_j$ is the jitter. Uniform priors were adopted for
all the model parameters.

\begin{figure}[h]
\begin{center}
\includegraphics[width=0.95 \columnwidth]{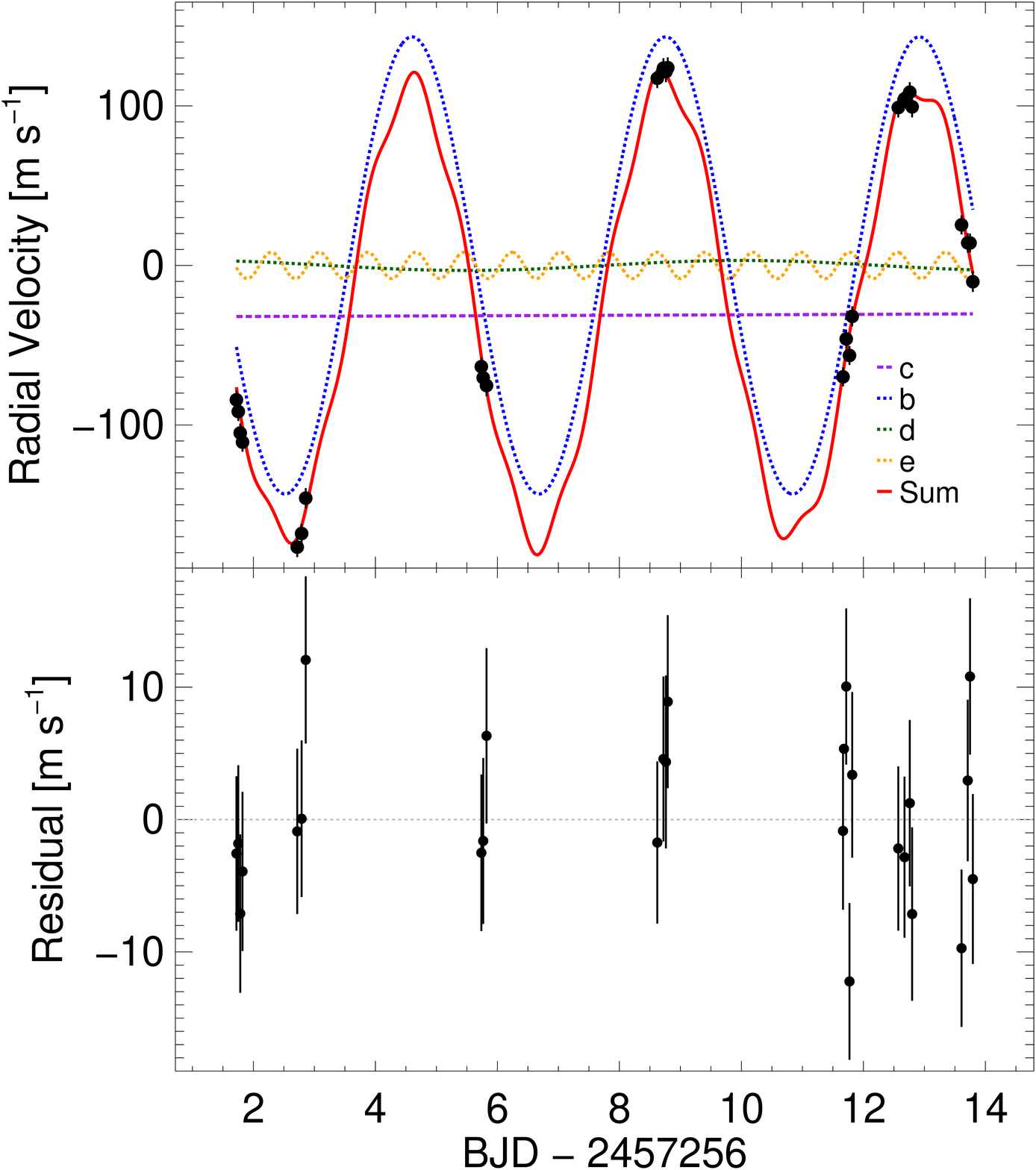}
\caption{{\it Top.}---Measured radial velocity of WASP-47 (black dots)
  along with the best-fitting model (red line).  The contributions to
  the model from each planet are also plotted as colored curves. {\it
    Bottom.}---Differences between the data and the best-fitting
  model.  The error bars represent the quadrature sum of the
  internally-estimated measurement uncertainties, and the fitted
  jitter parameter ($6.1^{+1.4}_{-1.1}$~m~s$^{-1}$).}
\label{rvall}
\vspace{-0.1cm}
\end{center}
\end{figure}

\begin{figure}[h]
\begin{center}
\includegraphics[width = 1.0\columnwidth]{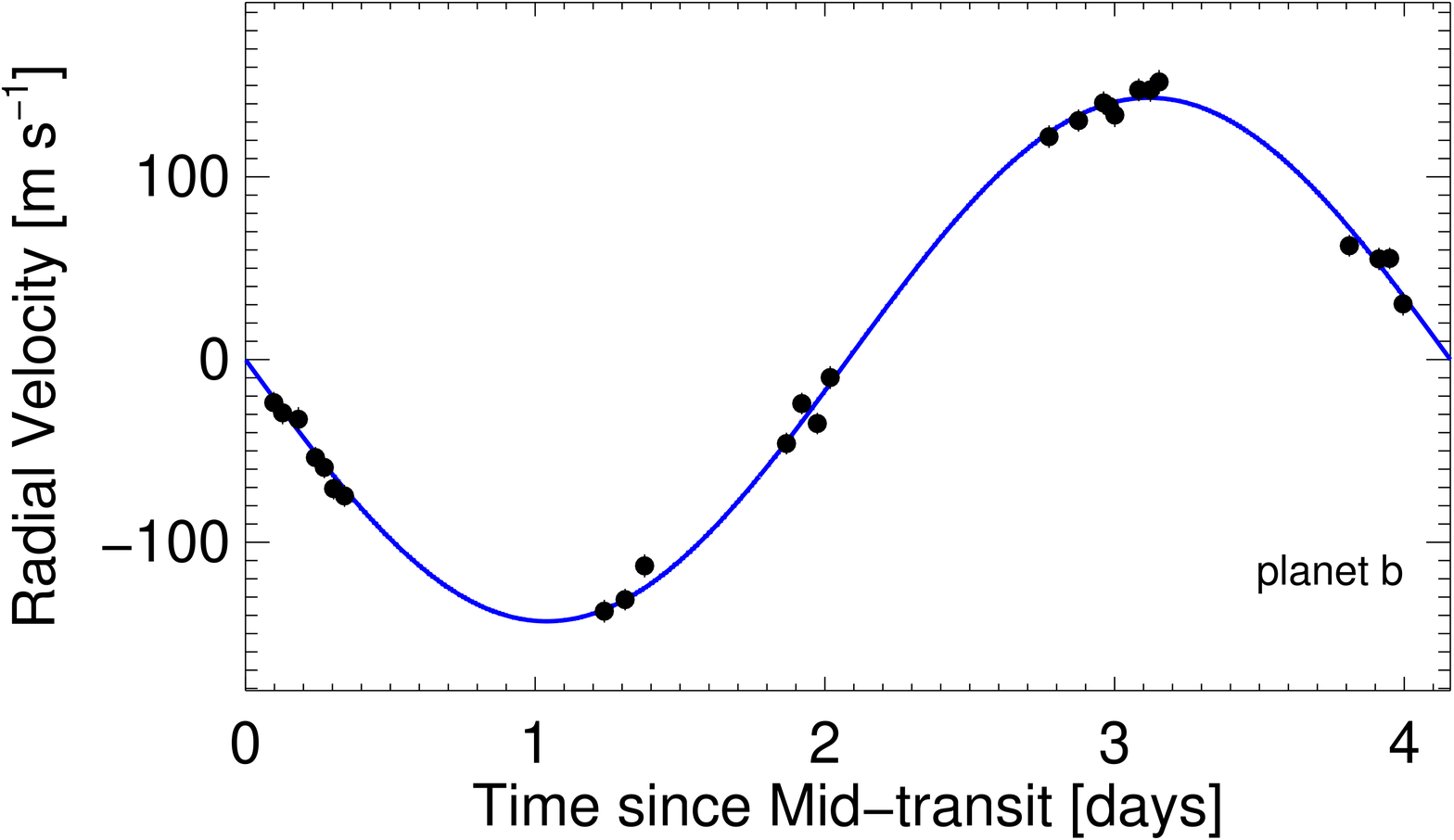}
\includegraphics[width = 1.0\columnwidth]{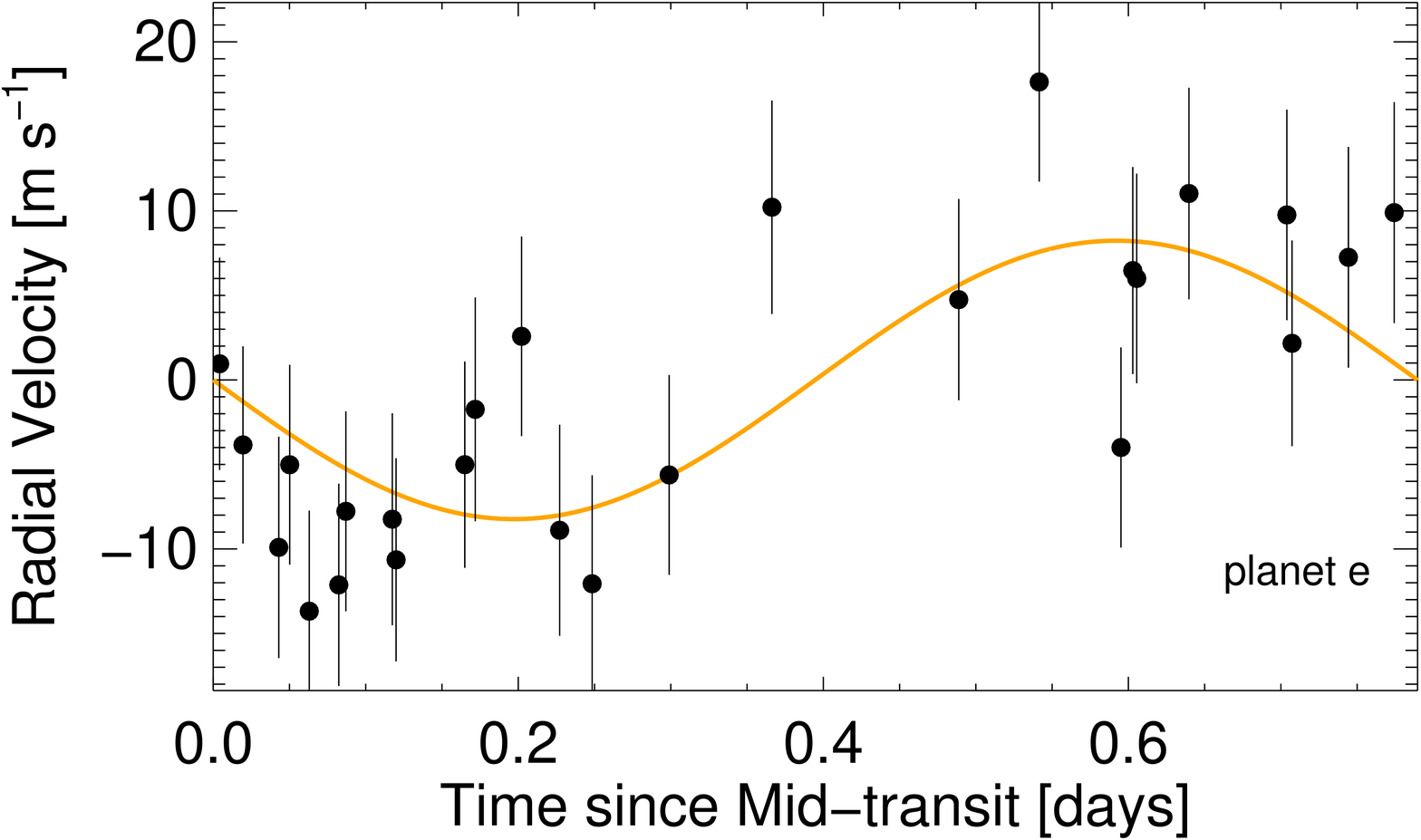}
\includegraphics[width = 1.0\columnwidth]{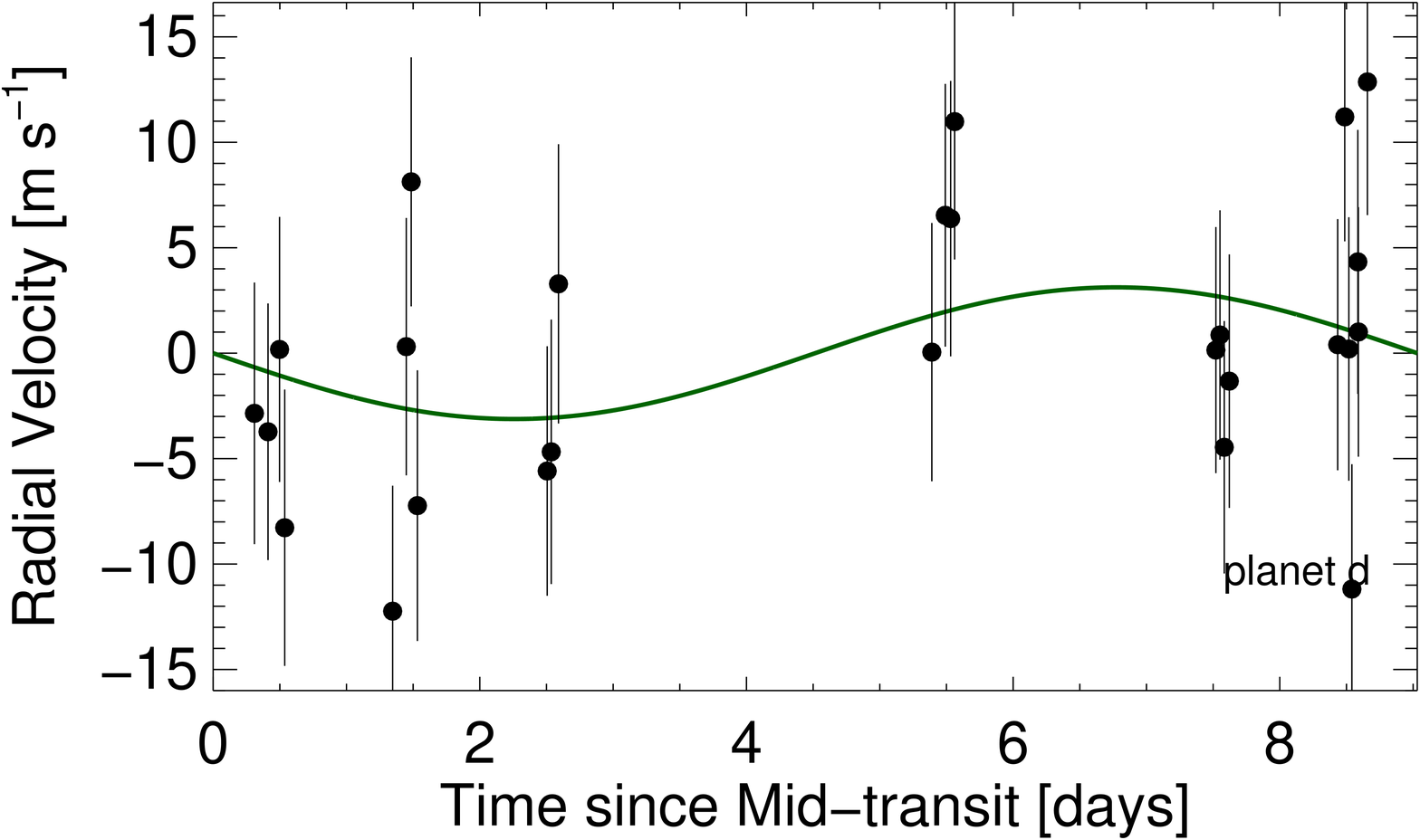}
\end{center}
\caption{Radial velocity as a function of the orbital phase of each of
  the three inner planets. In each case, the modeled contributions of
  the other two planets has been removed, before plotting.}
\label{fold}
\end{figure}

We maximized the likelihood using the Nelder-Mead (``Amoeba'') method.
The best-fitting model is shown by a red line in the top panel of
Figure~\ref{rvall}. The lower panel shows the residuals.  Figure
\ref{fold} shows the radial-velocity variation specific to each
planet, based on the data and the parameters of the best-fitting
model.

To determine the parameter uncertainties and covariances we employed a
Markov Chain Monte Carlo method. In particular we used the
affine-invariant ensemble sampler proposed by \citet{Goodman2010}. We
started 100 chains in a Gaussian ``ball'' in the neighborhood of the
best-fitting model parameters. We stopped the chains when the
Gelman-Rubin potential scale reduction factor \citep{Gelman1992}
dropped to 1.01, a standard criterion for adequate convergence. The
posterior probabilities of all parameters are smooth and
unimodal. Table \ref{tbl:params} reports the results. The reported
``best fit'' value is the median of the marginalized posterior
distribution.  The reported uncertainty interval encompasses the range
between the 16$\%$ and 84$\%$ percentile levels of the cumulative
distribution. The result for the jitter parameter was
$6.1^{+1.4}_{-1.1}$~m~s$^{-1}$, about twice as large as the internal
measurement uncertainty.

The motion induced by the hot Jupiter WASP-47b was clearly detected,
with a semi-amplitude $K_b = 143.3 \pm 2.5$~m~s$^{-1}$. This result is
consistent with the previously reported Doppler data of
\citet{Hellier}, who found $K_b = 136 \pm 5$~m~s$^{-1}$.  The inner
planet WASP-47e was detected at the $3.4\sigma$ level, with $K_c = 8.2
\pm 2.4$~m~s$^{-1}$.  The motion induced by the outer planet WASP-47d
was detected weakly if at all, leading to the result $ K_d = 3.1 \pm
2.5$ ~m~s $^{-1}$ if we allow the semi-amplitude to range over both
positive and negative numbers. If we require the semi-amplitude to be
positive, we obtain an upper limit of 8.3~m~s$^{-1}$ with 95\%
confidence.

The root-mean-squared residual between the data and
the best-fitting three-planet model was 6.08~m~s$^{-1}$.  As another
measure of the significance of each planet detection, we tried
refitting the data with different numbers of planets.  When only
planet b is modeled (i.e., $K_c = K_d \equiv 0$) the root-mean-squared
(rms) residual between the data and the best-fitting model is
9.16~m~s$^{-1}$.  When WASP-47e is included in the model, the rms
residual drops to 6.34~m~s$^{-1}$. It drops further to 6.08~m~s$^{-1}$
when all three planets are modeled. 

The planetary masses can be calculated from the semi-amplitudes,
orbital periods, and stellar mass. In doing so we adopted a stellar
mass of $M_\star = 1.04\pm 0.08~M_\odot$ \citep{mortier}. The results
are given in Table \ref{tbl:params}a. This table also provides
planetary mean densities, based on the masses from our work and the
radii reported by \citet{Becker} based on the {\it K2} transit
photometry. The best-fitting model satisfies the heuristic criterion
for long-term dynamical stability that was proposed by
\citet{fabrycky2014}: $\Delta_{\rm in} + \Delta_{\rm out}> 18 $, where
$\Delta$ is the difference between semi-major axes of the inner and
outer pairs of planets measured in terms of their mutual Hill radius.

\begin{figure}[h]
\begin{center}

\includegraphics[width = 0.95\columnwidth]{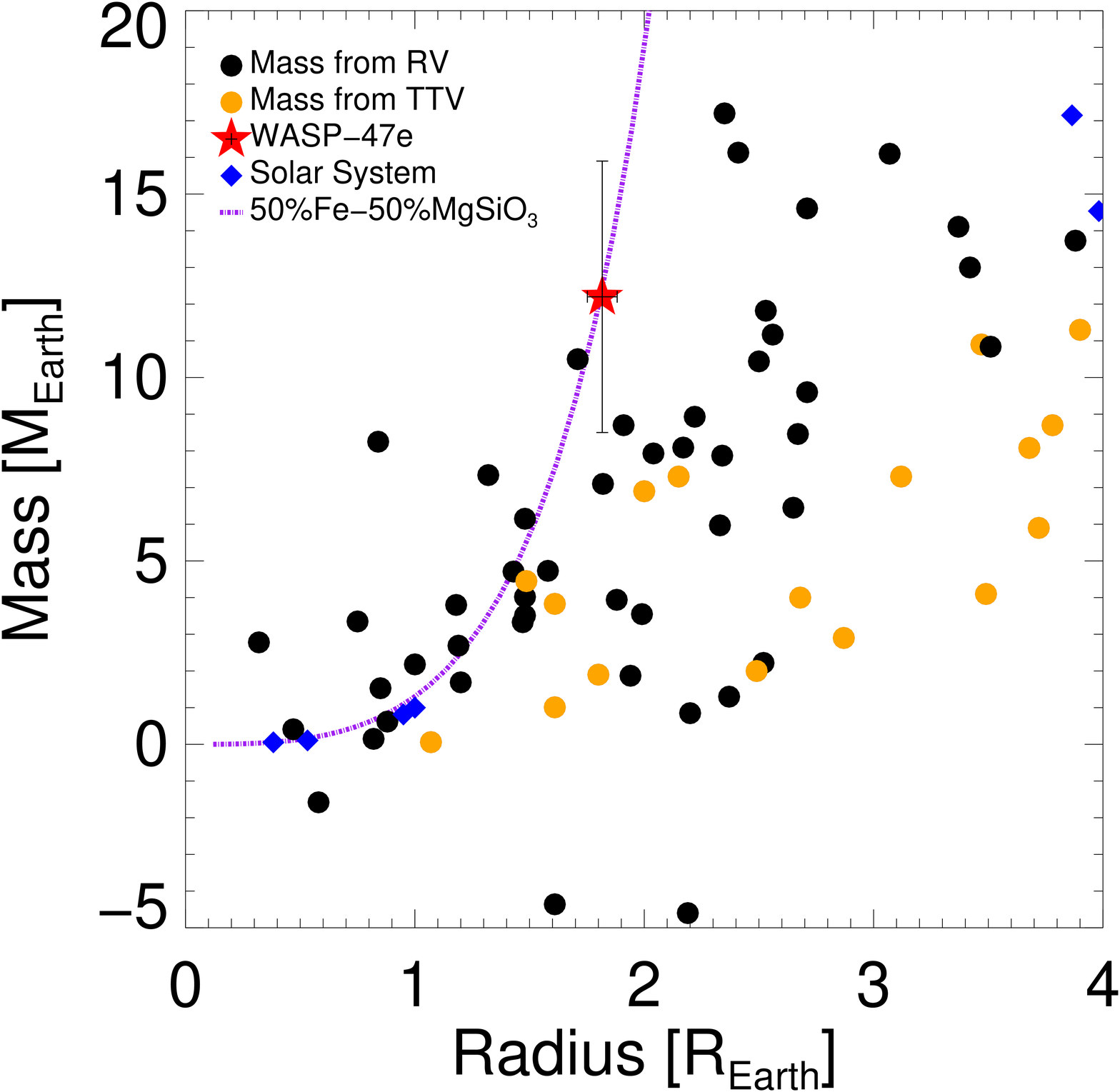}
\includegraphics[width = 0.95\columnwidth]{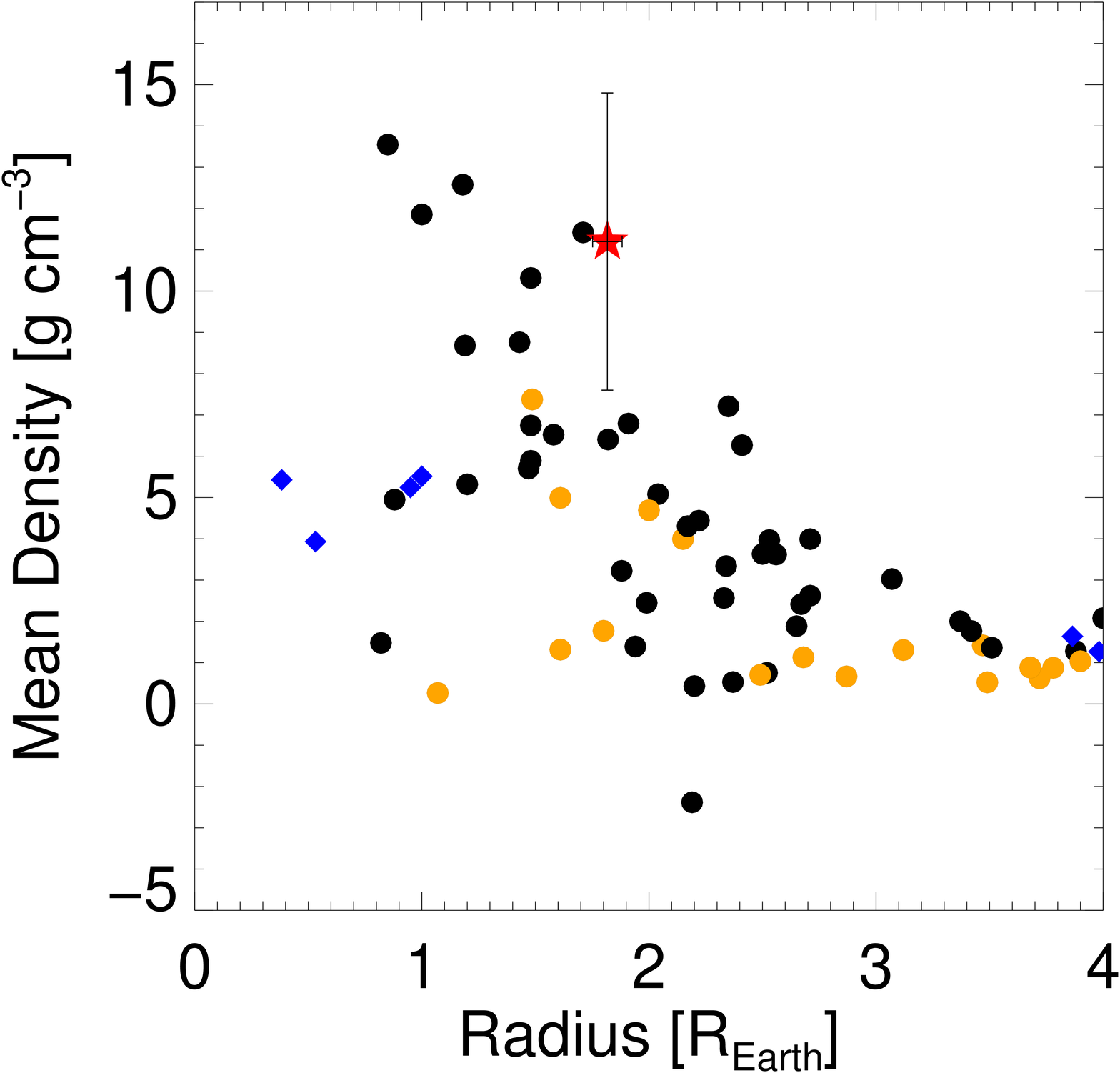}
\end{center}
\caption{{\it Top.}---Masses and sizes of exoplanets
  \citep[compiled by][see references therein]{Wolfgang}, as measured
  through the Doppler method (black) and the TTV method
  (orange). The blue diamonds are solar
  system planets. The red star is WASP-47e, which lies close to the theoretical
  curve for a composition of 50\%~Fe and 50\% MgSiO$_{3}$ \citep{zeng}
  (a stony-iron composition). {\it Bottom.}---Mean densities and sizes
  of the same sample.}
\label{rm}
\end{figure}

\section{Discussion}
\label{sec:discussion}

The $3.4\sigma$ detection of the radial-velocity variation induced by
WASP-47e further diminishes the probability that the
corresponding transit signal is an ``astrophysical false positive''
due to an unresolved eclipsing binary. Thus, WASP-47 is unambiguously
the first known case of a hot Jupiter accompanied by a shorter-period,
smaller planetary companion. The compactness and apparent flatness of
the system, along with the prograde rotation of the host star
\citep{RM2015}, seem consistent with quiescent migration
through the protoplanetary disk \citep{Lin1996}, as opposed to
dynamically hotter scenarios involving planet-planet encounters or
perturbations from distant stars. Simulations by \citet{Mustill} showed that high-eccentricity migration of a giant planet is often destructive to the inner planetary system.

Barring subsequent disruption, neighboring planets undergoing disk migration are likely to be trapped in first-order mean-motion resonance (MMR) \citep[e.g.][]{peale_1976}. It is therefore interesting
that the period ratios between the inner three planets of WASP-47 are
not especially close to any first-order MMR.  The period ratio between
the outer pair (b and d) is 2.17, which is quite typical of the period
ratios for neighboring planets observed in the {\it Kepler}
multiplanet systems \citep{SteffenHwang2015}. There does not yet seem
to be a convincing explanation for the prevalence of this period
ratio.  Furthermore, the ratio of 5.27 for the inner two planets (b
and e) conforms with a trend noted by \citet{SteffenFarr2013}: the
shortest-period planets ($\lesssim 2$~days) tend to have unusually
large period ratios with their closest neighbors.

With a period shorter than one day, the innermost planet e is an
example of an ``ultra-short period'' (USP) planet, as classified by
\citet{usp}. Those authors found that USP planets are almost always
smaller than 2~$R_\oplus$ and frequently appear in compact
multiple-planet systems, and indeed WASP-47e has both of these
characteristics. Relatively few masses have been measured for USP
planets. They are generally expected to be rocky, with thin or
nonexistent atmospheres, because of the intense heat from the nearby
star and the possibility of photoevaporation of any thick atmosphere.
One notable exception is 55~Cnc~e, an USP planet that does seem to
have a thick atmosphere, judging from its relatively low mean density
\citep{Winn2011,Demory2011}. In contrast, our mass and radius
measurements for WASP-47e imply a mean density of $11.2 \pm
3.6$~g~cm$^{-3}$, which is consistent with a rocky planet. For
example, the measured dimensions of WASP-47e are compatible with the
models of \citet{zeng} for a stony-iron composition (50\% Fe and 50\%
MgSiO$_{3}$), as illustrated in Figure~\ref{rm}.

In fact the measured mass and density of WASP-47e are both somewhat
higher than would be predicted based on the few previous measurements
of planets in the same size range. The empirical mass-radius
relationship of \citet{weissmarcy2014} predicts a mass for WASP-47e of
4.5~$M_\oplus$.  Likewise, the probabilistic relationship of
\citet{Wolfgang} predicts a mass of $5.9\pm 1.9~M_\oplus$, and gives a
likelihood of only $\sim$3\% for a planet with the measured dimensions
of WASP-47e. \citet{Rogers} argued that $1.6~R_{\oplus}$ represents a
critical radius, separating smaller planets with a mainly rocky
composition from larger planets with substantial low-density
atmospheres. WASP-47e with its radius of $1.8~R_{\oplus}$ and
apparently rocky composition, seems to be an exception to this
rule. This discrepancy may be because the planet samples analyzed by
\citet{Rogers} were not as strongly irradiated as WASP-47e. The strong
irradiation could have stripped the planet of its volatile atmosphere,
thereby leaving behind the dense rocky core. Based on current
estimates of the stellar parameters and the orbital distance of
WASP-47e, the planet receives roughly 3800 times more stellar
radiation than the Earth. The combination of strong irradiation and large radius of WASP-47e is in agreement with the finding of
\citet{wolfganglopez} who showed that the inclusion of the flux
dependence broadens the radius range over which the expected
composition changes from rocky to volatile-enhanced.  In their
models, the transition ranges over 1.2--1.8~$R_{\oplus}$.

It is also interesting to compare the Doppler and TTV methods for
measuring the planet masses.  For the giant planet, the TTV mass of
$341^{+73}_{-55}~M_{\oplus}$ is within about 1$\sigma$ of our Doppler
mass of $370 \pm 29~M_{\oplus}$.  For the inner planet, the TTV
analysis led to an upper limit of 22~$M_\oplus$, which is compatible
with our Doppler mass of $12.2 \pm 3.7~M_{\oplus}$. For the outer
planet, the Doppler mass constraint of $10.4\pm 8.4~M_{\oplus}$ is
consistent with the TTV constraint of
$15.2^{+6.7}_{-7.6}~M_{\oplus}$. All these results are in accordance,
although there is still plenty of scope for improving both
measurements and sharpening the comparison, particularly for the
smaller planets.

Another intriguing feature of WASP-47 is the diversity among the
properties of the inner three planets. The radius ratios, $R_{b}/R_{e}
= 7.0 \pm 0.5$ and $R_{b}/R_{e} = 3.5 \pm 0.2$, are among the most
extreme of neighboring planets observed by {\it Kepler}. Out of the
1020 neighboring pairs that appear in the NASA Exoplanet Archive
catalog of Kepler Objects of Interest (KOIs), only 27 have a radius
ratio larger than $3.5$, and only 2 pairs have radius ratio larger
than $7.0$. The density contrast between b and e is $11 \pm 5$, which
is subject to large uncertainty but may be even higher than the
density contrast of $8.4 \pm 1.5$ between Kepler-36b and c
\citep{Carter}, an exemplar of the phenomenon of dissimilar
neighboring planets. Curiously, the progression of the apparent
compositions of planets e, b, and d, from rocky to Jovian to
volatile-enhanced, matches the order we see in our much more
spread-out solar system.

The discovery and confirmation of additional close planetary
companions of this previously known hot Jupiter also raises the
question: how many of the other known hot Jupiters have close-in
companions?  A picture had been developing that hot Jupiters are
``lonely'' in the sense of lacking close planetary companions
\citep{Steffen2012,Wright}, but have we really excluded the possibility of
small USP planets among the sample of hundreds of known hot Jupiters?
It seems worthwhile to scrutinize those systems in greater detail,
through ground-based photometry, space-based photometry with the {\it
  K2} and upcoming {\it TESS} missions \citep{TESS}, and perhaps even
with intensive Doppler programs.

\acknowledgements We thank the referee for a very stimulating review
of the manuscript. Work by FD and JNW was supported by the NASA
Origins program (grant NNX11AG85G) as well as the {\it Transiting
  Exoplanet Survey Satellite} mission. AV is supported by the National
Science Foundation Graduate Research Fellowship, Grant No. DGE
1144152. Australian access to the Magellan Telescopes was supported by
the Australian Federal Government through the Department of Industry
and Science.

{\it Facilities:} \facility{Magellan:Clay (Planet Finder
  Spectrograph)}


\begin{thebibliography}{}
\expandafter\ifx\csname natexlab\endcsname\relax\def\natexlab#1{#1}\fi

\bibitem[{{Agol} {et~al.}(2005){Agol}, {Steffen}, {Sari}, \&
  {Clarkson}}]{Agol2005}
{Agol}, E., {Steffen}, J., {Sari}, R., \& {Clarkson}, W. 2005, \mnras, 359, 567

\bibitem[{{Becker} {et~al.}(2015){Becker}, {Vanderburg}, {Adams}, {Rappaport},
  \& {Schwengeler}}]{Becker}
{Becker}, J.~C., {Vanderburg}, A., {Adams}, F.~C., {Rappaport}, S.~A., \&
  {Schwengeler}, H.~M. 2015, ArXiv e-prints, arXiv:1508.02411

\bibitem[{{Butler} {et~al.}(1996){Butler}, {Marcy}, {Williams}, {McCarthy},
  {Dosanjh}, \& {Vogt}}]{Butler}
{Butler}, R.~P., {Marcy}, G.~W., {Williams}, E., {et~al.} 1996, \pasp, 108, 500

\bibitem[{{Carter} {et~al.}(2012){Carter}, {Agol}, {Chaplin}, {Basu},
  {Bedding}, {Buchhave}, {Christensen-Dalsgaard}, {Deck}, {Elsworth},
  {Fabrycky}, {Ford}, {Fortney}, {Hale}, {Handberg}, {Hekker}, {Holman},
  {Huber}, {Karoff}, {Kawaler}, {Kjeldsen}, {Lissauer}, {Lopez}, {Lund},
  {Lundkvist}, {Metcalfe}, {Miglio}, {Rogers}, {Stello}, {Borucki}, {Bryson},
  {Christiansen}, {Cochran}, {Geary}, {Gilliland}, {Haas}, {Hall}, {Howard},
  {Jenkins}, {Klaus}, {Koch}, {Latham}, {MacQueen}, {Sasselov}, {Steffen},
  {Twicken}, \& {Winn}}]{Carter}
{Carter}, J.~A., {Agol}, E., {Chaplin}, W.~J., {et~al.} 2012, Science, 337, 556

\bibitem[{{Crane} {et~al.}(2010){Crane}, {Shectman}, {Butler}, {Thompson},
  {Birk}, {Jones}, \& {Burley}}]{crane}
{Crane}, J.~D., {Shectman}, S.~A., {Butler}, R.~P., {et~al.} 2010, in Society
  of Photo-Optical Instrumentation Engineers (SPIE) Conference Series, Vol.
  7735, Society of Photo-Optical Instrumentation Engineers (SPIE) Conference
  Series, 53

\bibitem[{{Demory} {et~al.}(2011){Demory}, {Gillon}, {Deming}, {Valencia},
  {Seager}, {Benneke}, {Lovis}, {Cubillos}, {Harrington}, {Stevenson}, {Mayor},
  {Pepe}, {Queloz}, {S{\'e}gransan}, \& {Udry}}]{Demory2011}
{Demory}, B.-O., {Gillon}, M., {Deming}, D., {et~al.} 2011, \aap, 533, A114

\bibitem[{{Dressing} {et~al.}(2015){Dressing}, {Charbonneau}, {Dumusque},
  {Gettel}, {Pepe}, {Collier Cameron}, {Latham}, {Molinari}, {Udry}, {Affer},
  {Bonomo}, {Buchhave}, {Cosentino}, {Figueira}, {Fiorenzano}, {Harutyunyan},
  {Haywood}, {Johnson}, {Lopez-Morales}, {Lovis}, {Malavolta}, {Mayor},
  {Micela}, {Motalebi}, {Nascimbeni}, {Phillips}, {Piotto}, {Pollacco},
  {Queloz}, {Rice}, {Sasselov}, {S{\'e}gransan}, {Sozzetti}, {Szentgyorgyi}, \&
  {Watson}}]{Dressing2015}
{Dressing}, C.~D., {Charbonneau}, D., {Dumusque}, X., {et~al.} 2015, \apj, 800,
  135

\bibitem[{{Fabrycky} {et~al.}(2014){Fabrycky}, {Lissauer}, {Ragozzine}, {Rowe},
  {Steffen}, {Agol}, {Barclay}, {Batalha}, {Borucki}, {Ciardi}, {Ford},
  {Gautier}, {Geary}, {Holman}, {Jenkins}, {Li}, {Morehead}, {Morris},
  {Shporer}, {Smith}, {Still}, \& {Van Cleve}}]{fabrycky2014}
{Fabrycky}, D.~C., {Lissauer}, J.~J., {Ragozzine}, D., {et~al.} 2014, \apj,
  790, 146
\bibitem[Goodman~\&\ Weare(2010)]{Goodman2010} Goodman,~J. \& Weare,\ J., 2010, Comm.\ App.\ Math.\ Comp.\ Sci., 5, 65
\bibitem[Gelman~\&\ Rubin(1992)]{Gelman1992} Gelman,~A. \& Rubin,\ D., 1992, Stat.\ Sci., 7, 457-472

\bibitem[{{Hellier} {et~al.}(2012){Hellier}, {Anderson}, {Collier Cameron},
  {Doyle}, {Fumel}, {Gillon}, {Jehin}, {Lendl}, {Maxted}, {Pepe}, {Pollacco},
  {Queloz}, {S{\'e}gransan}, {Smalley}, {Smith}, {Southworth}, {Triaud},
  {Udry}, \& {West}}]{Hellier}
{Hellier}, C., {Anderson}, D.~R., {Collier Cameron}, A., {et~al.} 2012, \mnras,
  426, 739

\bibitem[{{Holman} {et~al.}(1997){Holman}, {Touma}, \& {Tremaine}}]{Holman1997}
{Holman}, M., {Touma}, J., \& {Tremaine}, S. 1997, \nat, 386, 254

\bibitem[{{Holman} \& {Murray}(2005)}]{Holmanmurray}
{Holman}, M.~J., \& {Murray}, N.~W. 2005, Science, 307, 1288

\bibitem[{{Howard} {et~al.}(2013){Howard}, {Sanchis-Ojeda}, {Marcy}, {Johnson},
  {Winn}, {Isaacson}, {Fischer}, {Fulton}, {Sinukoff}, \&
  {Fortney}}]{Howard2013}
{Howard}, A.~W., {Sanchis-Ojeda}, R., {Marcy}, G.~W., {et~al.} 2013, \nat, 503,
  381

\bibitem[{{Howell} {et~al.}(2014){Howell}, {Sobeck}, {Haas}, {Still},
  {Barclay}, {Mullally}, {Troeltzsch}, {Aigrain}, {Bryson}, {Caldwell},
  {Chaplin}, {Cochran}, {Huber}, {Marcy}, {Miglio}, {Najita}, {Smith},
  {Twicken}, \& {Fortney}}]{howell}
{Howell}, S.~B., {Sobeck}, C., {Haas}, M., {et~al.} 2014, \pasp, 126, 398

\bibitem[{{Innanen} {et~al.}(1997){Innanen}, {Zheng}, {Mikkola}, \&
  {Valtonen}}]{Innanen}
{Innanen}, K.~A., {Zheng}, J.~Q., {Mikkola}, S., \& {Valtonen}, M.~J. 1997,
  \aj, 113, 1915

\bibitem[{{Jontof-Hutter} {et~al.}(2014){Jontof-Hutter}, {Lissauer}, {Rowe}, \&
  {Fabrycky}}]{Jontof2014}
{Jontof-Hutter}, D., {Lissauer}, J.~J., {Rowe}, J.~F., \& {Fabrycky}, D.~C.
  2014, \apj, 785, 15

\bibitem[{{Jontof-Hutter} {et~al.}(2015){Jontof-Hutter}, {Rowe}, {Lissauer},
  {Fabrycky}, \& {Ford}}]{Jontof2015}
{Jontof-Hutter}, D., {Rowe}, J.~F., {Lissauer}, J.~J., {Fabrycky}, D.~C., \&
  {Ford}, E.~B. 2015, \nat, 522, 321

\bibitem[{{Lin} {et~al.}(1996){Lin}, {Bodenheimer}, \& {Richardson}}]{Lin1996}
{Lin}, D.~N.~C., {Bodenheimer}, P., \& {Richardson}, D.~C. 1996, \nat, 380, 606

\bibitem[{{Lissauer} {et~al.}(2013){Lissauer}, {Jontof-Hutter}, {Rowe},
  {Fabrycky}, {Lopez}, {Agol}, {Marcy}, {Deck}, {Fischer}, {Fortney}, {Howell},
  {Isaacson}, {Jenkins}, {Kolbl}, {Sasselov}, {Short}, \&
  {Welsh}}]{Lissauer2013}
{Lissauer}, J.~J., {Jontof-Hutter}, D., {Rowe}, J.~F., {et~al.} 2013, \apj,
  770, 131

\bibitem[{{Marcy} {et~al.}(2014){Marcy}, {Isaacson}, {Howard}, {Rowe},
  {Jenkins}, {Bryson}, {Latham}, {Howell}, {Gautier}, {Batalha}, {Rogers},
  {Ciardi}, {Fischer}, {Gilliland}, {Kjeldsen}, {Christensen-Dalsgaard},
  {Huber}, {Chaplin}, {Basu}, {Buchhave}, {Quinn}, {Borucki}, {Koch}, {Hunter},
  {Caldwell}, {Van Cleve}, {Kolbl}, {Weiss}, {Petigura}, {Seager}, {Morton},
  {Johnson}, {Ballard}, {Burke}, {Cochran}, {Endl}, {MacQueen}, {Everett},
  {Lissauer}, {Ford}, {Torres}, {Fressin}, {Brown}, {Steffen}, {Charbonneau},
  {Basri}, {Sasselov}, {Winn}, {Sanchis-Ojeda}, {Christiansen}, {Adams},
  {Henze}, {Dupree}, {Fabrycky}, {Fortney}, {Tarter}, {Holman}, {Tenenbaum},
  {Shporer}, {Lucas}, {Welsh}, {Orosz}, {Bedding}, {Campante}, {Davies},
  {Elsworth}, {Handberg}, {Hekker}, {Karoff}, {Kawaler}, {Lund}, {Lundkvist},
  {Metcalfe}, {Miglio}, {Silva Aguirre}, {Stello}, {White}, {Boss}, {Devore},
  {Gould}, {Prsa}, {Agol}, {Barclay}, {Coughlin}, {Brugamyer}, {Mullally},
  {Quintana}, {Still}, {Thompson}, {Morrison}, {Twicken}, {D{\'e}sert},
  {Carter}, {Crepp}, {H{\'e}brard}, {Santerne}, {Moutou}, {Sobeck}, {Hudgins},
  {Haas}, {Robertson}, {Lillo-Box}, \& {Barrado}}]{marcy2014}
{Marcy}, G.~W., {Isaacson}, H., {Howard}, A.~W., {et~al.} 2014, \apjs, 210, 20

\bibitem[{{Mazeh} {et~al.}(1997){Mazeh}, {Krymolowski}, \& {Rosenfeld}}]{Mazeh}
{Mazeh}, T., {Krymolowski}, Y., \& {Rosenfeld}, G. 1997, \apjl, 477, L103

\bibitem[{{Meschiari} {et~al.}(2009){Meschiari}, {Wolf}, {Rivera}, {Laughlin},
  {Vogt}, \& {Butler}}]{systemic}
{Meschiari}, S., {Wolf}, A.~S., {Rivera}, E., {et~al.} 2009, \pasp, 121, 1016

\bibitem[{{Mortier} {et~al.}(2013){Mortier}, {Santos}, {Sousa}, {Fernandes},
  {Adibekyan}, {Delgado Mena}, {Montalto}, \& {Israelian}}]{mortier}
{Mortier}, A., {Santos}, N.~C., {Sousa}, S.~G., {et~al.} 2013, \aap, 558, A106

\bibitem[{{Mustill} {et~al.}(2015){Mustill}, {Davies}, \& {Johansen}}]{Mustill}
{Mustill}, A.~J., {Davies}, M.~B., \& {Johansen}, A. 2015, \apj, 808, 14

\bibitem[{{Neveu-VanMalle} {et~al.}(2015){Neveu-VanMalle}, {Queloz},
  {Anderson}, {Brown}, {Collier Cameron}, {Delrez}, {D{\'{\i}}az}, {Gillon},
  {Hellier}, {Jehin}, {Lister}, {Pepe}, {Rojo}, {S{\'e}gransan}, {Triaud},
  {Turner}, \& {Udry}}]{Neveu}
{Neveu-VanMalle}, M., {Queloz}, D., {Anderson}, D.~R., {et~al.} 2015, ArXiv
  e-prints, arXiv:1509.07750

\bibitem[{{Peale}(1976)}]{peale_1976}
{Peale}, S.~J. 1976, \araa, 14, 215

\bibitem[{{Pepe} {et~al.}(2013){Pepe}, {Cameron}, {Latham}, {Molinari}, {Udry},
  {Bonomo}, {Buchhave}, {Charbonneau}, {Cosentino}, {Dressing}, {Dumusque},
  {Figueira}, {Fiorenzano}, {Gettel}, {Harutyunyan}, {Haywood}, {Horne},
  {Lopez-Morales}, {Lovis}, {Malavolta}, {Mayor}, {Micela}, {Motalebi},
  {Nascimbeni}, {Phillips}, {Piotto}, {Pollacco}, {Queloz}, {Rice}, {Sasselov},
  {S{\'e}gransan}, {Sozzetti}, {Szentgyorgyi}, \& {Watson}}]{Pepe2013}
{Pepe}, F., {Cameron}, A.~C., {Latham}, D.~W., {et~al.} 2013, \nat, 503, 377

\bibitem[{{Rasio} \& {Ford}(1996)}]{Rasio}
{Rasio}, F.~A., \& {Ford}, E.~B. 1996, Science, 274, 954

\bibitem[{{Ricker} {et~al.}(2014){Ricker}, {Winn}, {Vanderspek}, {Latham},
  {Bakos}, {Bean}, {Berta-Thompson}, {Brown}, {Buchhave}, {Butler}, {Butler},
  {Chaplin}, {Charbonneau}, {Christensen-Dalsgaard}, {Clampin}, {Deming},
  {Doty}, {De Lee}, {Dressing}, {Dunham}, {Endl}, {Fressin}, {Ge}, {Henning},
  {Holman}, {Howard}, {Ida}, {Jenkins}, {Jernigan}, {Johnson}, {Kaltenegger},
  {Kawai}, {Kjeldsen}, {Laughlin}, {Levine}, {Lin}, {Lissauer}, {MacQueen},
  {Marcy}, {McCullough}, {Morton}, {Narita}, {Paegert}, {Palle}, {Pepe},
  {Pepper}, {Quirrenbach}, {Rinehart}, {Sasselov}, {Sato}, {Seager},
  {Sozzetti}, {Stassun}, {Sullivan}, {Szentgyorgyi}, {Torres}, {Udry}, \&
  {Villasenor}}]{TESS}
{Ricker}, G.~R., {Winn}, J.~N., {Vanderspek}, R., {et~al.} 2014, in Society of
  Photo-Optical Instrumentation Engineers (SPIE) Conference Series, Vol. 9143,
  Society of Photo-Optical Instrumentation Engineers (SPIE) Conference Series,
  20

\bibitem[{{Rogers}(2015)}]{Rogers}
{Rogers}, L.~A. 2015, \apj, 801, 41

\bibitem[{{Sanchis-Ojeda} {et~al.}(2014){Sanchis-Ojeda}, {Rappaport}, {Winn},
  {Kotson}, {Levine}, \& {El Mellah}}]{usp}
{Sanchis-Ojeda}, R., {Rappaport}, S., {Winn}, J.~N., {et~al.} 2014, \apj, 787,
  47

\bibitem[{{Sanchis-Ojeda} {et~al.}(2015){Sanchis-Ojeda}, {Winn}, {Dai},
  {Howard}, {Isaacson}, {Marcy}, {Petigura}, {Sinukoff}, {Weiss}, {Albrecht},
  {Hirano}, \& {Rogers}}]{RM2015}
{Sanchis-Ojeda}, R., {Winn}, J.~N., {Dai}, F., {et~al.} 2015, ArXiv e-prints,
  arXiv:1509.05337

\bibitem[{{Schmitt} {et~al.}(2014){Schmitt}, {Agol}, {Deck}, {Rogers}, {Gazak},
  {Fischer}, {Wang}, {Holman}, {Jek}, {Margossian}, {Omohundro}, {Winarski},
  {Brewer}, {Giguere}, {Lintott}, {Lynn}, {Parrish}, {Schawinski}, {Schwamb},
  {Simpson}, \& {Smith}}]{Schmitt}
{Schmitt}, J.~R., {Agol}, E., {Deck}, K.~M., {et~al.} 2014, \apj, 795, 167

\bibitem[{{Steffen} \& {Farr}(2013)}]{SteffenFarr2013}
{Steffen}, J.~H., \& {Farr}, W.~M. 2013, \apjl, 774, L12

\bibitem[{{Steffen} \& {Hwang}(2015)}]{SteffenHwang2015}
{Steffen}, J.~H., \& {Hwang}, J.~A. 2015, \mnras, 448, 1956

\bibitem[{{Steffen} {et~al.}(2012){Steffen}, {Ragozzine}, {Fabrycky}, {Carter},
  {Ford}, {Holman}, {Rowe}, {Welsh}, {Borucki}, {Boss}, {Ciardi}, \&
  {Quinn}}]{Steffen2012}
{Steffen}, J.~H., {Ragozzine}, D., {Fabrycky}, D.~C., {et~al.} 2012,
  Proceedings of the National Academy of Science, 109, 7982

\bibitem[{{Weidenschilling} \& {Marzari}(1996)}]{Weidenschilling}
{Weidenschilling}, S.~J., \& {Marzari}, F. 1996, \nat, 384, 619

\bibitem[{{Weiss} \& {Marcy}(2014)}]{weissmarcy2014}
{Weiss}, L.~M., \& {Marcy}, G.~W. 2014, \apjl, 783, L6

\bibitem[{{Winn} {et~al.}(2011){Winn}, {Matthews}, {Dawson}, {Fabrycky},
  {Holman}, {Kallinger}, {Kuschnig}, {Sasselov}, {Dragomir}, {Guenther},
  {Moffat}, {Rowe}, {Rucinski}, \& {Weiss}}]{Winn2011}
{Winn}, J.~N., {Matthews}, J.~M., {Dawson}, R.~I., {et~al.} 2011, \apjl, 737,
  L18

\bibitem[{{Wolfgang} \& {Lopez}(2015)}]{wolfganglopez}
{Wolfgang}, A., \& {Lopez}, E. 2015, \apj, 806, 183

\bibitem[{{Wolfgang} {et~al.}(2015){Wolfgang}, {Rogers}, \& {Ford}}]{Wolfgang}
{Wolfgang}, A., {Rogers}, L.~A., \& {Ford}, E.~B. 2015, ArXiv e-prints,
  arXiv:1504.07557

\bibitem[{{Wright} {et~al.}(2009){Wright}, {Upadhyay}, {Marcy}, {Fischer},
  {Ford}, \& {Johnson}}]{Wright}
{Wright}, J.~T., {Upadhyay}, S., {Marcy}, G.~W., {et~al.} 2009, \apj, 693, 1084

\bibitem[{{Zeng} \& {Sasselov}(2013)}]{zeng}
{Zeng}, L., \& {Sasselov}, D. 2013, \pasp, 125, 227

\end{thebibliography}
\end{document}